# Classification of White Blood Cells Using Machine and Deep Learning Models: A Systematic Review

Rabia Asghar, Sanjay Kumar, Paul Hynds, Arslan Shaukat

Rabia Asghar is Lecturer in Department of Computer Engineering at University of Lahore (UoL), Pakistan (Email: rabia.asghar@ dce.uol.edu.pk)
Corresponding Author : rabia.asghar@ dce.uol.edu.pk
Sanjay Kumar is doing PhD in Department of Electronic and Computer Engineering at University of Limerick Ireland in the domain of computer vision and machine learning.
(Email: chowdhry.sanjay@yahoo.com)
Paul Hynds is senior research fellow in the Environmental Sustainability and Health Institute (ESHI) at Dublin Institute of Technology, Ireland. His work employs statistical, mathematical and machine learning tools to predict the presence, frequency, and movement of contaminants in the Irish environment, and how, when and where these contaminants cause adverse human health effects. (Email: paul.hynds@tudublin.ie )
Arslan Shaukat is Associate Professor in Machine Learning & Head PG Program at National University of Sciences and Technology (NUST), Pakistan
 (Email: arslan.shaukat@ce.ceme.edu.pk)

*Abstract*— **Machine learning (ML) and deep learning (DL) models have been employed to significantly improve analyses of medical imagery, with these approaches used to enhance accuracy of prediction and classification. Model predictions and classifications assist diagnoses of various cancers and tumors. This review presents an in-depth analysis of modern techniques applied within the domain of medical image analysis for white blood cell classification. The methodologies that use blood smear images, magnetic resonance imaging (MRI), X-rays, and similar medical imaging domains are identified and discussed, with a detailed analysis of ML/DL techniques applied to the classification of white blood cells (WBCs) representing the primary focus of the review. The data utilized in this research has been extracted from a collection of 136 primary papers that were published between the years 2006 and 2023. The most widely used techniques and best-performing white blood cell classification methods are identified. While the use of ML and DL for white blood cell classification has concurrently increased and improved in recent year, significant challenges remain - 1) Availability of appropriate datasets remain the primary challenge, and may be resolved using data augmentation techniques. 2) Medical training of researchers is recommended to improve current understanding of white blood cell structure and subsequent selection of appropriate classification models. 3) Advanced DL networks including Generative Adversarial Networks, R-CNN, Fast R-CNN, and faster R-CNN will likely be increasingly employed to supplement or replace current techniques.**

## I. INTRODUCTION

White blood cells play a vital role in the human body's immune system. They identify and neutralize pathogens including bacteria, viruses, and cancer cells. Classification of WBCs is vital for accurate and early diagnosis and treatment of a range of diseases and medical conditions [1]. Machine learning techniques, both traditional and deep, have been widely adopted for myriad applications, including medical image analysis (MIA). MIA is critical in modern healthcare systems, aiding medical professionals in making well-informed decisions. It is used to diagnose various diseases, including brain tumors, lung cancer, anemia, leukemia, and malaria, via a range of image modalities including Magnetic Resonance Imaging (MRI), Computed Tomography (CT-Scans), Ultrasounds, Positron Emission Tomography (PET), Blood Smear images, and hybrid modalities [2]. Within MIA, these imageries are essential for analyzing and classifying hard and soft tissues in human organs for diagnostic and research purposes [3]. Accordingly, MIA has attracted significant attention from computer vision experts, with traditional and deep machine learning techniques having been applied in leukocyte segmentation, cancer detection, classification, medical image annotation, and image retrieval in computer-aided diagnosis (CAD). The efficacy of these methods therefore directly influences the clinical diagnosis and treatment strategies, highlighting the significance of technological advancements, such as high-speed computational resources and improved hardware and storage capabilities, for CAD [4-5]. One of the primary application areas for CAD systems using traditional machine learning and deep learning is the segmentation and classification of leukocytes (white blood cells). Leukocytes provide valuable information to medical professionals (doctors, hematologists, pathologists, and radiologists), for diagnosing various blood-related issues, including Human Immunodeficiency Virus (HIV) and blood cancer (leukemia). Changes in the count of white blood cells (WBCs) and/or morphological cell alterations, for instance variations in size, shape, and color observed in blood smear images, can provide valuable insights into various health disorders [6-9].

The blood cells are categorized into three major types: white blood cells (leukocytes), red blood cells (erythrocytes), and platelets (thrombocytes). Leukocytes are subdivided into five types: monocytes, lymphocytes, neutrophils, basophils, and eosinophils (Figure 1). Over the past two decades, significant advances have been made in traditional ML and DL methods for classification and segmentation of WBCs in microscopic blood smear images. Conventional methods depend on manually analyzing these images using microscope, which is tedious and error prone. Accordingly, the development of automated and computer-aided systems has become crucial in accurate, systematic, unbiased and rapid clinical diagnosis and effective treatment. Automated analysis of WBCs in blood smear images can significantly reduce the workload of hematologists and provide fast, accurate, and efficient results to assist medical professionals in the diagnostic process [10-13]. There are two overarching methods typically used to achieve automated WBC classification in blood smear images: traditional machine learning (ML) and deep learning (DL) techniques. These techniques have the potential to make medical hematology more efficient. The generalized overview of machine learning and deep learning techniques used to classify WBCs is depicted in Figure 2. Various computer-aided systems can automatically diagnose different types of hematic diseases, for example HIV and blood cancer, by analyzing leukocytes. The traditional machine learning process involves interconnected steps such as segmenting the region of interest and extracting the features, followed by optimal classification [14-15]. The feature extraction phase in traditional machine learning methods is challenging and directly impacts the classification performance. More recently, deep learning approaches are increasingly used due to =higher performance and decreasing complexity. Advanced deep learning methods with transfer learning have further improved implementation of automated systems for classification of WBCs. This study explores ML and DL techniques for classifying white blood cells in blood smear images. While deep learning and machine learning are explored in medical image analysis (MIA), a gap remains in white blood cell classification. This paper addresses this gap by comprehensively analyzing ML and DL methods, focusing on classifying five white blood cell types. The research highlights a significant interest in classifying white blood cells, mainly in image classification and segmentation. Some researchers use handcrafted features in their methods. Initial approaches involve data preprocessing, feature selection, and extraction [16]. A current approach is the use of deep learning to enhance white blood cell classification.

- This paper identifies and provides a thorough analysis of ML and DL models to classify white blood cells.
- Performance of ML and DL methods are discussed, and their applications in clinical practice are investigated. For the future work the challenges and limitations of the models are discussed.

Section II presents the review methodology used in this study. In Section III, the dataset is explained in detail. Section IV discusses the significance of medical imaging techniques, divided into two subsections: Traditional Machine Learning and Deep Learning. Moving on, Section V reviews and highlights the limitations of previous studies and discusses the challenges faced. Section VI outlines the future directions for research in this field. Lastly, overarching conclusions are presented in Section VII.

## II. REVIEW METHODOLOGY

The research done throughout the years to classify white blood cells is shown in Figure 3. The traditional machine learning (ML) and deep learning (DL) methodologies are discussed separately because there is clear evidence of the difference between these two methods. The researchers started from the conventional methods but, with time, shifted towards the deep learning models due to their promising performance. Deep learning methodologies have grown exponentially recently, specifically in leucocyte classification in blood smear images.

A well-organized and planned review process was essential to identify, scan, include/exclude and synthesize targeted literature which satisfies the search criteria and uses existing resources [16]. In the current review, the authors sought to incorporate the most advanced and related research articles based on manual and automatic searches to identify all significant content. The approach was initiated by identifying pertinent research questions. The two research questions (RQ) formulated in accordance with the PICO (Population, Intervention, Comparison, Outcome) search framework are as follows:

(i) *How have systems been developed for the grouping of WBCs based on ML and DL?*

(ii) *What are the applications of traditional machine learning and deep learning methods for effectively classifying WBCs in blood smear images?*

The keywords were extracted from RQs, as represented in Table 1.

**Table 1:** Keywords for searching databases

| Blood Cells | | | |
|---|---|---|---|
| Leukocyte(s) classification | A1 | White blood cell(s) detection | A2 |
| White blood cell(s) classification | A3 | White blood cell(s) segmentation | A4 |
| **Machine Learning** | | | |
| Machine learning | B1 | Deep learning | B2 |
| Big data | B3 | Artificial Intelligence | B4 |

The next phase after RQ development was identification of relevant articles/studies via automated searching of electronic databases based on extracted keywords from RQs. The recent articles published from 2006 to May 2023 are included in the proposed survey. To align with the study's emphasis on recent research trends and technological progress, articles prior to 2006 were omitted. Research articles were searched in significant repositories such as Google Scholar, Scopus, and Web of Science. The inclusion and exclusion criteria depicted in Table 2.

**Table 2:** Inclusion and Exclusion Criteria for Proposed Research

| Inclusion Criteria | Exclusion Criteria |
|---|---|
| Only English-language articles are included in the survey. | Articles other than English language articles are excluded. |
| Relevant articles to the topic are incorporated in the survey. | Irrelevant articles are excluded. |
| The articles having the clear results and methodology are included in the survey. | The articles having the poor results and methodology are excluded in the survey. |
| Period: 2016 to May 2023. | Period: Pre 2016 |

During the review on the classification of leukocytes, a total of 136 research articles were collected, and an in-depth quality assessment of articles were performed to select the best papers. Quality assessment of a research paper is critical in assessing a research article's consistency, validity, and overall credibility [18]. The quality assessment was performed by considering various aspects such as topic relevancy, the methodology and machine learning models used, dataset collection and data analysis, clarity and consistency of the results presented in the articles, journal reputation, and highly cited papers.

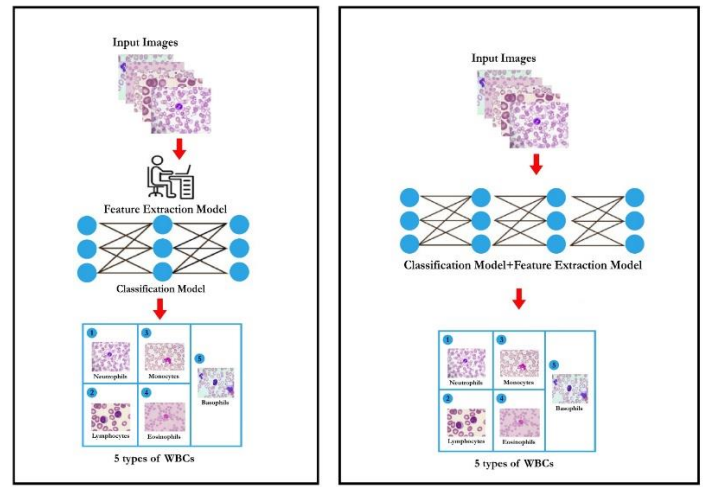

(a)          (b)

**Figure 2:** Neutrophils classification in blood images (a) Traditional machine learning model (b) Advanced deep learning model.

### III. DATASETS

Different authors utilized a variety of sources for white blood cell datasets, resulting in a wide range of dataset sizes for training, testing, and validation. In total, 27 studies were identified, incorporating datasets like ALL-IDB, Private Dataset, CellaVision, AA-IDB2, Hayatabad Medical Dataset, Isfahan Al-Zahra and Omid hospital, ALL-IDB2/Leishman stained peripheral blood smears, Public Dataset, BCCD, Kaggle, LISC and BCCD, Jiangxi Tecom Science Corporation/CellaVision/Bsisc/LISC, KMC hospital Manipal India, Hybrid-Leukocyte database/e Hybrid-Slide database, Acquired from Sixth People's Hospital of Shenzhen, SMC-IDB/IUMS-IDB/ALL-IDB, and SBILab. The list of datasets along with the corresponding number of images is provided in Table 3.

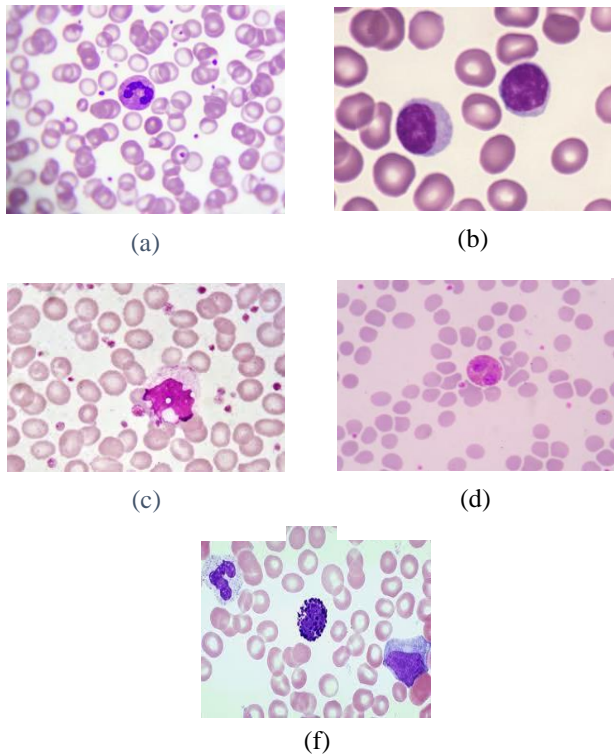

**Figure 1:** Categories of white blood cells [17] (a) Neutrophils (b) Lymphocytes (c) Monocytes (d) Eosinophils (e) Basophils

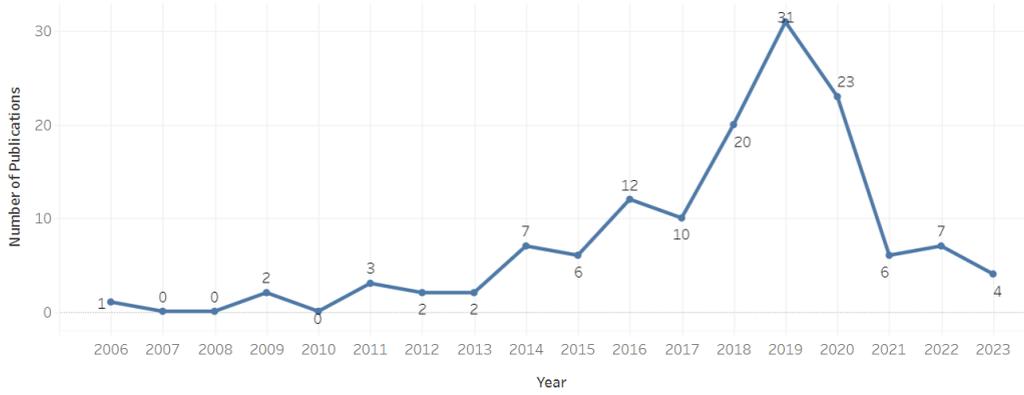

Figure 3: The number of publications remained relatively constant from 2006 to 2013. Starting from 2014, there was a gradual increase in the number of publications, reaching its peak in 2019. However, there was a significant decrease in publications after 2019, mainly due to the COVID-19 pandemic and its impact on data accessibility.

Table 3: White blood cell datasets

| Dataset | No of Images |
|---|---|
| ALL-IDB [58] | 130 |
| Private Dataset [61] | 70 |
| CellaVision [66] | 100 |
| AA-IDB2 [67] | 108 |
| Hayatabad Medical Dataset [69] | 1030 |
| Isfahan Al-Zahra and Omid hospital [70] | 312 |
| Private Dataset [73] | 431 |
| ALL-IDB2/ Leishman stained peripheral blood smears [78] | 160/160 |
| CellaVision [80] | 450 |
| Public Dataset [85] | 92,800 |
| BCCD [86] | 12,444 |
| Kaggle [87] | 12,444 |
| BCCD [89] | 12,500 |
| BCCD [91] | 375 |
| Kaggle/LISC [92] | 12,500/400 |
| LISC and BCCD [94] | 6250 |
| Jiangxi Tecom Science Corporation/ CellaVision/ Bsisc/ LISC [95] | 300/ 100/ 268/ 257 |
| KMC hospital, Manipal, India [96] | 280 |
| ALL-IDB[101] | 108 |
| Hybrid-Leukocyte database/ e Hybrid-Slide database [102] | 891/ 377 |
| Acquired from Sixth People's Hospital of Shenzhen [104] | 21 |
| Kaggle [105] | 12,494 |
| SMC-IDB/ IUMS-IDB/ ALL-IDB [108] | 367/ 195/ 108 |
| BCCD [110] | 12447 |
| SBILab [113] | 76 |
| BCCD [117] | 2487 |
| Kaggle [119] | 12,444 |

## IV. SIGNIFICANCE OF MEDICAL IMAGING TECHNIQUE

As mentioned above, medical imaging includes many modalities including X-rays, CT scans, MRIs, and microscopic blood smear images. These images are vital to allow medical practitioners detect diseases via MIA. Most researchers have used three parameters for performance evaluation: accuracy, sensitivity, and specificity. Accuracy represents the proportion of correct predictions, sensitivity shows the rate of true positives (TP) classified by the model, while specificity shows the percentage of true negatives (TN) accurately identified by the system. These parameters are computed by using equations (1-3).

$$\text{Accuracy} = \frac{TP+TN}{TP+TN+FP+FN} \quad (1)$$

$$\text{Sensitivity} = \frac{TP}{TP+FN} \quad (2)$$

$$\text{Specificity} = \frac{TN}{TN+FP} \quad (3)$$

TP and TN represent correctly classified true positive and true negative samples, while FP and FN show incorrectly classified false positive and false negative instances.

### A. White Blood Cell Classification Using Traditional Machine Learning

Various studies have explored traditional machine learning methods for white blood cell (WBC) classification, organized into preprocessing-based techniques, feature extraction, and classification. A total of 39 studies were identified, wherein 13 papers (33.33%) focused on pre-processing techniques, 15 papers (38.46%) delved into feature extraction methods, and 11 papers (28.21%) emphasized classification techniques for white blood cell classification. This distribution is evident in tables 4 to 7, highlighting diverse emphases on these sub-processes within the conventional machine learning approach for classifying white blood cells.

#### I. Preprocessing-based ML techniques

Preprocessing-based techniques include methods that manipulate and enhance raw data before further analysis. In the context of white blood cell (WBC) classification, these techniques play a critical role in refining images to enable accurate categorization. Rosyadi et al. [19] used optical microscopy to generate blood samples images. The method consists of four stages: image pre-processing, segmentation, feature extraction, and classification. In the first phase of image preprocessing, images were transformed from RGB to grayscale and binary images. Subsequently, in the second phase, resizing, cropping, and edge detection were applied to all images. Five geometrical features were considered in the feature extraction phase that represent important geometric characteristics of the segmented cells: normalized area, solidity, eccentricity, circularity, and normalized perimeter. These characteristics help differentiate various types of white blood cells and enable accurate classification through K-means clustering. The study focus was analysis of each feature for accuracy. After experimentation, it was concluded that the circularity feature was most significant as it achieved 67% accuracy, which was the highest. The eccentricity feature had the lowest accuracy of up to 43%.

Gautam et al. [20] also presented a technique that started with preprocessing of microscopic images. Preprocessing involved conversion of RGB images to grayscale, contrast stretching, and histogram equalization. Subsequently, they applied segmentation through Otsu's thresholding method, followed by geometrical feature extraction, including perimeter, area, eccentricity, and circularity. Finally, a Naïve bayes classifier was used for classification with the maximum likelihood method, achieving an accuracy of 80.88% of white blood cells classification.

Savkare et al. [21] presented an alternative method for segmenting blood cells. The process started with preprocessing, using median and Laplacian filters to enhance image quality. After preprocessing, images were transformed from RGB (Red, Green, Blue) to HSV (Hue, Saturation, Value) color space. Then K-mean clustering was applied for segmentation of blood cells. Furthermore, they used morphological operation and a watershed algorithm to refine cell separation. The proposed method through K-mean clustering produced satisfactory results with an accuracy of 95.5%.

#### II. ML Techniques with a focus on feature extraction

Typically, a differential counting method of white blood cells is used to assess the patient's immune system. This method involves using flow cytometry and fluorescent markers, which can disturb the cell due to repetitive sample preparation. Accordingly, label-free techniques that uses imaging flow cytometry and ML algorithms to classify unstained WBCs are considered a more effective approach. Toh et al. [22] previously reported impressive average F1-score of 97% across B and T subtypes, with each individual subtype achieving a distinct F1 score of 78%.

Tsai et al. [23] proposed a multi-class support vector machine (SVM) to hierarchically identify and categorize blood cell images. During this process, segmentation was implemented on digital images to retrieve geometric features from each segment, enabling identification and classification of different blood cells types. The experiment demonstrated that utilizing hierarchical multi-class SVM classification led to significant improvements. The experimental outcomes were compared with manual results, revealing that the proposed method outperformed manual classification with a notable accuracy of 95.3%. Likewise, Şengür et al. [24] presented a model combining image processing (IP) and machine learning (ML) techniques for classification of

WBCs Shape-based features and deep features were utilized to describe the WBCs. For classification, a long-short-term memory (LSTM) model was applied, with a dataset comprising 349 blood images with 10-fold cross-validation, from which 35 geometric and statistical features were extracted for training and testing purposes. Six ML techniques (decision tree classifier, Random Forest, K-Nearest Neighbor, Multinomial Logistic Regression, Naïve Bayes, and SVM) were used to categorize white blood cells [25]. Using shape-based features, an accuracy of 80% was achieved, while deep features achieved 82.9% accuracy. When both were used, 85.7% accuracy was obtained. To assess the algorithm's performance, five types of data were prepared for both training and testing purposes, resulting in 100 different combinations for each dataset. Findings showed that Multinomial Logistic Regression (MLR) returned the highest precision rate of 95%, Followed by Random Forest. The authors concluded that performance could be further enhanced via bagging, bootstrapping, or boosting.

Huang et al. [26] presented a technique for white blood cell segmentation, delineating their approach into several phases: nucleus segmentation and recognition, feature extraction, and classification. In nucleus segmentation, a leukocyte nucleus enhancer (LNE) was used to enhance the contrast of nucleus colors. After LNE implementation, multiple levels of Otsu's thresholding were applied, effectively preserving only the leukocytes and suppressing other cell types. During the feature extraction phase, a gray-level co-occurrence matrix was employed in which 80 texture features were extracted. Subsequently, they incorporated shape-based features, including compactness and roughness, after which Principal Component Analysis (PCA) was used to reduce feature dimensions. Classification was achieved using a genetic-based parameter selector (GBPS) with 50-time cross-validation The study's findings revealed a remarkable accuracy of 95%, validating that the proposed genetic algorithm outperformed K-means clustering for WBC classification.

Yampri et al. [27] also used blood imagery as input and segmented out the WBC via automatic thresholding (i.e., segregate the nucleus and cytoplasm of cells) and feature extraction. Eigen cells were used to remove parts by applying the following steps: conversion of cell image to vector, computation of mean and covariance of vector, computation of eigen values and eigen vectors. Principle component analysis (PCA) was used to transform high dimensional eigen space to significantly lower dimensional space, with 92% accuracy achieved.

## III. Techniques with a focus on classification

Tavakoli et al [28] developed a three-phase ML method for improved WBC classification of WBCs, with the process divided as follows: find the nuclei and cytoplasm, extract features, and classification. A novel process was designed to segment the whole nucleus - segmentation of cytoplasm involves location detection inside the convex region. In the next phase, four unique color and three shape features were extracted, and finally, in the last phase, SVM was used for WBC classification. Study results showed 94.2% accuracy with the BCCD data set, 92.21% with LISC data, and 94.65% with Raabin datasets of WBCs, however, only Raabin-WBC datasets were not associated with hyperparameter issues.

An innovative "Computer-aided diagnosis system" method was proposed to diagnose blood from microscopic images, with this process utilizing a hybrid approach, whereby CNN was employed as a feature extractor. The performance of these classifiers was measured and with Random Forest (RF) outperforming other classifiers based on a 98.7% accuracy [29].

Gupta et al. [30] present an optimized form of the Binary Bat algorithm inspired by bat echolocation techniques. Using OBBA (Table 3), dimensionality reduction was achieved by eliminating ≥11 similar features. Subsequently four classifiers (KNN, Logistic Regression, RF, and DT) were applied to the dataset for WBC classification. OBBA demonstrated remarkable performance, achieving an average accuracy of 97.3% surpassing other optimizers like the Optimized Crow Search Algorithm (OCSA), which attained an accuracy of 92.77% and the Optimized Cuttlefish Algorithm (OCFA), with an accuracy of 95.17%. Lee [31] proposed an innovative technique of image segmentation based on grey-level thresholding. It was found that the cell-type specific reaction of the cells produces adequate evidence to allow precise classification. This method was tested on a dataset comprising 1149 white blood cells from 13 altered, clinically significant categories. Cells were randomly selected from 20 blood smears obtained from leukemia patients. Sorting these cells on the base of quantitative amounts in the segmented images produced an accuracy of 82.6%.

### B. Deep Learning Techniques

Wibawa et al. [32] proposed a model for classifying two WBC types, comparing the results with three conventional machine learning methods (support vector machines), using nine features for classification. The authors found that deep learning significantly surpasseed conventional ML methods, achieving high accuracy of 95.5%.

Toğaçar [33] introduced a WBC classification approach based on the coefficient and Ridge feature selection method utilizing a CNN model with GoogleNet and ResNet50 for feature extraction. They achieved 97.95% accuracy for classification and counting of WBCs. A CNN was employed to identify and classify each segmented WBC image as being granular or non-granular. Subsequently, granular cells were further categorized into eosinophils and neutrophils, while non-granular cells were classified as lymphocytes and monocytes [34]. To enhance the dataset's robustness, augmentation approaches were implemented, resulting in improved accuracy for both binary and multiclassification of blood cell subtypes, equating to 98.51% precision for WBC classification and 97.7% precision for subtypes classification.

Lippeveld et al. [35] examined four human blood samples using Image flow cytometry. Two models were used to identify eight WBC types, while the other two samples were for identifying eosinophils. ML models were applied to both datasets to classify human blood cells with 5-fold cross validation. Random Forest (RF) and Gradient Boosting (GB) were used for the first model, while deep learning CNN architecture (ResNet and DeepFlow (DF)) were employed for the second model. On the white blood cell dataset, results demonstrate a relatively balanced accuracy of 0.778 and 0.70, while similarly for the eosinophil dataset, a balanced accuracy of 0.871 and 0.856 was achieved. The GB and RF classifiers demonstrated similar performance, with GB exhibiting higher accuracy on the EOS dataset. For the WBC dataset, precisions were 0.774 and 0.778, respectively, while for the EOS dataset, they achieved 0.825 and 0.871. Additionally, the DF outperformed the RN architecture on the WBC dataset, attaining a balanced accuracy of 0.703 compared to RN's 0.649.

Rawat et al. [36] introduced another deep learning method employing the DenseNet121 model for classification of numerous types of WBC - The proposed model was estimated, with an accuracy of 98.84%. Results indicate that the DenseNet121 model with a batch size of 8 exhibited highest overall performance. The dataset, consisting of 12,444 images, was obtained from Kaggle. Nazlibilek et al. [37] proposed a DL-based method that leveraged image variation operations and generative adversarial networks (GAN) for accurately classifying white blood cells into five distinct types. Likewise, Sadeghian et al. [38] developed a two-stage model comprising an initial alteration using a pre-trained model, followed by the integration of a ML classifier. They employed the BCCD dataset, a downscaled blood cell detection dataset, and achieved a precision of 97.03%. Likewise, Sadeghian et al. [38] developed a two-stage model comprising an initial alteration using a pre-trained model, followed by the integration of a ML classifier. They employed the BCCD dataset, a downscaled blood cell detection dataset, and achieved a precision of 97.03%.

Macawile et al. [39], utilized Convolutional Neural Networks (CNNs) to effectively classify and count white blood cells (WBCs) in microscopic blood images. Among the proposed models AlexNet, GoogleNet, and ResNet-101. AlexNet performed better than the other two. It demonstrated an impressive 89.18% sensitivity, 97.85% specificity, and an overall accuracy of 96.63%.

G. Liang et al. [40] introduced an innovative approach that seamlessly merges convolutional neural networks (CNNs) with recurrent neural networks (RNNs). This fusion termed the CNN-RNN framework, enhances understanding of image content and structured feature learning. It enables end-to-end training for comprehensive medical image data analysis. They applied transfer learning, adapting pre-trained weight parameters from the ImageNet dataset for the CNN segment. Additionally, they integrated a customized loss function to expedite training and achieve precise weight parameter convergence. Experimental results underscore their proposed model's outstanding accuracy and efficiency, notably achieving an accuracy of 90.79%. This achievement surpasses other CNN architectures such as ResNet and Inception V3.

Sharma et al. [41] present a deep learning methodology utilizing convolutional neural networks. Their model achieves an impressive 96% accuracy for binary classification and 87% accuracy for multiclass classification. This progress enhances the entire process, increasing precision and efficiency while decreasing the need for manual intervention.

Togacar et al. [42] did their research on the classification and identification of white blood cells (WBCs) using an exceptionally efficient computer-aided automated approach. Utilizing Regional-Based Helixal Neural Networks, their research effectively classified and differentiated white blood cells. The achievement of an impressive 99.52% accuracy in WBC classification.

E. H. Mohamed [43] introduced an alternative method for the identification and classification of blood cells based on CNN. The study presented two distinct approaches for classifying white blood cells. In the initial approach, a CNN was employed with transfer learning, utilizing pre-trained weight parameters applied to the images. In contrast, the second approach utilized Support Vector Machines (SVM) for the classification process. The classification results demonstrated a remarkable 98.4% accuracy for CNN and 90.6% accuracy for SVM. The classification results of CNN are higher compared to SVM.

M. Toğaçar et al. [44] introduced a method composed of three essential phases. In the initial stage, CNN models specifically AlexNet, GoogleNet, and ResNet-50 are utilized as feature extractors. Subsequently, the features extracted from these CNN model layers are fused. In the second phase, the technique incorporates feature selection methods, including MIC and Ridge Regression. In the third phase, these selected features are amalgamated. The overlapping features derived from the MIC and Ridge Regression techniques are then classified using the QDA method. This integrated approach achieves a remarkable overall success rate of 97.95% in classifying white blood cells.

X. Yao et al. [45] introduced a CNN-based approach for the classification of white blood cells. In their method, CNN integrated an optimizer to adaptively adjust parameters such as the learning rate, leveraging the efficient net architecture. The utilization of the optimizer responded to changes in loss and accuracy. Their proposed model demonstrated exceptional performance, achieving an impressive accuracy of 90%.

Khosrosereshki et al. [46] introduced an R-CNN-based model to identify the neutrophils, eosinophils, monocytes, and lymphocytes; two models were used. The first was Faster RCNN, and the second one was Yolov4. The study compared the classification accuracy of two models, Faster RCNN and Yolov4, and found that Faster RCNN obtained an accuracy of 96.25%. The accuracy of the one-stage model Yolov4 is 95%, while the Faster RCNN model exhibits superior performance over the Yolov4 model.

Bouchet et al. [47] introduced a deep-learning approach for the classification of white and red blood cells. The study utilized the Inception Recurrent Residual Convolutional Neural Network (IRRCNN) model, an advanced hybrid architecture based on residual networks and RCNN principles. The proposed IRRCNN demonstrated exceptional accuracy in experiments, achieving a 100% accuracy rate for classifying WBCs and 99.94% accuracy for RBCs.

K. K. Jha et al. [48] developed a leukemia detection module using deep learning, specifically designed for blood smear images. The detection process includes pre-processing, segmentation, feature extraction, and classification. The segmentation step utilizes a hybrid model based on Mutual Information (MI), which combines results from the active contour model and fuzzy C means algorithm. Subsequently, statistical and Local Directional Pattern (LDP) features are extracted from the segmented images. These features are then fed into a novel Deep CNN classifier based on the proposed Chronological Sine Cosine Algorithm (SCA) for classification purposes. The experimentation used blood smear images sourced from the AA-IDB2 database and evaluates performance using metrics such as True Positive Rate (TPR), True Negative Rate (TNR), and accuracy. The simulation results affirm that the Chronological SCA-based Deep CNN classifier achieves an impressive accuracy of 98.7%.

Ullah et al. [49] introduced a 3D-CNN feature-based CBVR system that is highly efficient and effective for retrieving similar content from vast video data repositories. After an in-depth exploration of its effectiveness in representing sequential frames, they selected middle layer features of a 3D-CNN model. Leveraging a mechanism for selecting convolutional features, only the active feature maps from the CNN layer that correspond to the ongoing event in the frame sequence are chosen. To condense the size of the extracted high-dimensional features for streamlined retrieval and expedited storage, they introduced the concept of hashing. These high-dimensional features are represented in compact binary codes through PCA, ensuring efficient search and reduced storage requirements for WBCs classification. For the classification of WBCs, the achieved accuracy is 85%.

S. Imran et al. [50] conducted a study involving the utilization of a four-hidden-layer feed-forward DNN and CNN. The research also extensively examines the impact of Mel-Frequency Cepstral Coefficients (MFCC) and Filter Bank Energies (FBE)features trained with various context sizes on two deep learning models, evaluated under normal, slow, and fast speaking rates. Micro-level analysis of results was conducted, revealing that the four-hidden-layer CNN slightly outperforms the DNN in classifying white blood cells. The CNN achieved an accuracy of 83% in classifying white blood cells. Notably, the experiments indicate that the CNN architecture, when coupled with FBE features, achieves improved performance across slow and fast speaking rates. This enhancement amounts to nearly 2% for fast and 3% for slow speaking rates.

Z. Kastrati et al. [51] introduced a convolutional neural network with three hidden layers, each having 1024 neurons, showcasing excellent performance in white blood cell classification on the INFUSE dataset, achieving a remarkable accuracy of 78.10%.

Ullah et al. [52] introduced an innovative conflux Long Short-Term Memory (LSTM) network for white blood cell classification. The framework involves four main stages: 1) frame-level feature extraction, 2) feature propagation through the conflux LSTM network 3) acquiring patterns and correlation computation, and 4) action classification. The process begins with extracting deep features using a pre-trained VGG19 CNN model from frame sequences for each view. Extracted features then undergo conflux LSTM processing to learn unique view-specific patterns. Interview correlations are computed by utilizing pairwise dot products from LSTM outputs across views, thus acquiring interdependent patterns. The VGG19 CNN model achieved an accuracy of 88.9%.

P. P. Banik et al. [53] presented an innovative convolutional neural network (CNN) model designed for WBC image classification. This model adeptly merges features from both the initial and final convolutional layers, while utilizing input image propagation through a convolutional layer to enhance performance. A dropout layer is added to counter overfitting. They achieved 98.61% average accuracy.

Y. Ku et al. [54] proposed an automated system for white blood cell classification using a dual-stage convolutional neural network (CNN). A dataset of 2,174 patch images was collected for training and testing purposes. The dual-stage CNN classifies images into 4 classes, achieving an accuracy of 97.06%, a precision of 97.13%, a recall of 97.06%, and an F-1 score of 97.1%.

M. Karthikeyan et al. [55] introduced the Leishman-stained function deep classification (LSM-TIDC) model for white blood cell (WBC) classification. The LSM-TIDC method explores the potential of interpolation and Leishman-stained function without the need for explicit segmentation. This approach effectively eliminates false regions in multiple input images. Following image preprocessing, relevant features are extracted through multi-directional feature extraction. A system is then developed, utilizing a transformation invariant model to extract nuclei and subsequently utilizing convolutional and pooling characteristics for cell classification. The effectiveness of the proposed method is verified through extensive experiments conducted on white blood cell images from Kaggle. The achieved accuracy was 94.42%.

A. Acevedo et al. [56] used a dataset of 17,092 peripheral blood cell images across eight classes that was gathered using the CellaVision DM96 analyzer. Pathologist-verified ground truth data was used for training two convolutional neural network architectures: Vgg-16 and Inceptionv3. In the first setup, networks acted as feature extractors for a support vector machine, achieving test accuracies of 86% (Vgg-16) and 90% (Inceptionv3). In the second setup, fine-tuning led to end-to-end models, yielding 96% accuracy (Vgg-16) and 95% accuracy (Inceptionv3).

Upon reviewing the studies presented in the identified literature, as a general observation, detection of WBCs through traditional methods (ML) tends to focus on segmenting cells after data pre-processing. These segmented data are then typically used for feature extraction in WBC classification. Accordingly, the traditional ML methods were associated with better results as accurate identification of white blood cells is impossible without segmenting resulting in higher levels of accuracy as shown in Tables 4-8. Research teams employed a range of methods for data segmentation and obtained a range of classification accuracies. While some traditional models achieved up to 99% accuracy, accuracy was shown to decrease in concurrence with dataset size. Deep learning models exhibited improved performance for more extensive datasets and exhibited higher levels of accuracy in concurrence with increasingly large datasets (Table 6). Some authors also implemented a combination of different datasets, in order to probe the accuracy of their models on unknown datasets (i.e., blind testing). Deep learning models have been a significant breakthrough in myriad domains and as shown in the identified literature, the use of traditional machine learning models within biomedical applications in general, and WBC classification in particular is undoubtedly shifting toward the use of deep learning models based on dataset size. But the deep learning models are now more mature and solving more complex problems with higher accuracy. There is a clear gap in using the latest advancements in deep learning, like transfer knowledge and the Meta-learning process. Apart from this, the authors focused on the same kind of datasets; one should collect new datasets based on the latest techniques and use them for their research.

Comparative analysis of deep learning models applied to different datasets revealed remarkably high levels of achieved accuracy across various studies (Table 9). Baghel et al. [113] demonstrated the high level of efficacy associated with the use of Convolutional Neural Networks (CNN), achieving an accuracy of 98.51%, while Riaz et al. [134] used a Convolutional Generative Adversarial Network (GAN) to obtain a classification accuracy of 99.9% on the Catholic University of Korea dataset. Mosabbir et al. [135] addressed the challenging National Institutes of Health (NIH) dataset using CNN and achieved remarkable results, attaining an accuracy of 97.92%. Tusneem et al. [136] also used CNN and demonstrated its strength, achieving 99.7% classification accuracy. Kakumani et al. [137] utilized a pre-trained InceptionV3 model on the Kaggle dataset and achieved 99.76% classification accuracy.

Popular WBC detection methods are recognized and presented in Table 4-8. The methods are divided into detection methodologies and combinations of ML and DL models. An analysis of ML and DL models was conducted, in which the DL approach outperform the ML approach.

Table 4: WBC nuclei detection accuracy, specificity and sensitivity in white blood smear images (n = 10)

| Author | Year | Method | Accuracy | Specificity | Sensitivity |
|---|---|---|---|---|---|
| Safuan et al. [57] | 2018 | Otsu thresholding and watershed marker | 98.87% | 96.87 | 99.10 |
| Huang et al. [58] | 2012 | Otsu thresholding | 98% | - | - |
| Danyali et al. [59] | 2015 | Fuzzy divergence threshold | 98% | | |
| Manik et al. [60] | 2016 | Adaptive thresholding | 98.9% | | |

| Author | Year | Method | Accuracy | Sensitivity | Specificity |
|---|---|---|---|---|---|
| Li et al. [61] | 2016 | Dual thresholding | 97.85% | | |
| Wang et al. [62] | 2106 | Spectral and morphologic | 90% | | |
| Negm et al. [63] | 2018 | K-mean clustering | 99.157% | 0.9952 | 0.99348 |
| Khosroseresliki et al. [64] | 2017 | Simple thresholding | 93.75% | - | - |
| Bouchet et al. [65] | 2019 | Fuzzy set algorithm | 99.32% | - | - |
| Jha et al. [66] | 2019 | Hybrid model based on Mutual Information | 98.7% | 0.98 | 0.98 |

**Table 5:** SVM accuracy, specificity and sensitivity for leucocyte classification (n = 4)

| Author | Year | Method | Accuracy | Sensitivity | Specificity |
|---|---|---|---|---|---|
| Duan et al. [67] | 2019 | SVM | 98.3% | - | - |
| Sajjad et al. [68] | 2016 | Support Vector Machine | 98.6% | 0.962 | 0.985 |
| Amin et al. [69] | 2015 | K-mean and SVM | 97% | 84.3 | 97.3 |
| Again et al. [70] | 2018 | Support Vector Machine | 94% | 0.9577 | 0.9787 |

**Table 6:** Different Various TML Models for leucocyte Classification (n = 6)

| Author | Year | Method | Accuracy | Sensitivity | Specificity |
|---|---|---|---|---|---|
| Gautam et al. [71] | 2016 | Naïve Bayesian classifier | 80.88% | - | - |
| Tantikitti et al. [72] | 2015 | Decision Tree | 92.2% | - | |
| Rawat et al. [73] | 2017 | PCA-SVM | 94.6% | 0.97 | 0.88 |
| Shaikhina et al. [74] | 2019 | Decision Tree and Random Forest classifier | 85% | 81.8 | 88.9 |
| Abdeldaim et al. [75] | 2018 | K-NN | 98.6% | - | - |
| Mathur et al. [76] | 2013 | Naïve Bayesian Classifier | 92.72% | 0.90 | - |

**Table 7:** ANN accuracy, specificity and sensitivity leucocyte classification (n = 7)

| Author | Year | Method | Accuracy | Sensitivity | Specificity |
|---|---|---|---|---|---|
| Hegde et al. [77] | 2019 | ANN | 99% | 99.4 | 0.9918 |
| Manik et al. [78] | 2016 | ANN | 98.9% | - | - |
| Su MC et al. [79] | 2014 | ANN | 99.11% | 0.973 | 0.982 |

| Author | Year | Method | Accuracy | Sensitivity | Specificity |
|---|---|---|---|---|---|
| Lee et al. [80] | 2014 | Hybrid Neural Network based Classifier | 91% | - | - |
| Rawat et al. [81] | 2018 | K-Means, ensemble artificial neural network (EANN) | 95% | - | - |
| Nazlibilek et al. [82] | 2014 | Neural Network Classifier | 95% | - | - |
| Sadeghian et al. [83] | 2009 | ANN | 78% | - | - |

**Table 8:** Deep learning model accuracy, specificity and sensitivity for leukocytes classification (n = 36)

| Author | Year | Method | Accuracy | Sensitivity | Specificity |
|---|---|---|---|---|---|
| M. J. Macawile et al. [84] | 2018 | AlexNet | 96.63% | 98.85% | 99.61% |
| G. Liang et al. [85] | 2018 | CNN + RNN | 91% | - | - |
| M. Sharma et al. [86] | 2019 | CNN | 97% | 94% | 98% |
| M. Togacar et al. [87] | 2019 | CNN | 97.78% | - | - |
| E.H. Mohamed et al. [88] | 2020 | Pre-trained Deep Learning Models | 97.03% | 0.71 | 0.91 |
| B. Ergen et al. [89] | 2020 | CNN, Feature Selection | 97.95% | 98 | 97.75 |
| C. Zhao et al. [90] | 2021 | TWO-DCNN | 96% | - | - |
| A. Cinar et al. [91] | 2021 | Alexnet-GoogleNet-SVM | 99.73%, 98.23% | 98.75 | - |
| Wang et al. [92] | 2019 | CNN Architecture SSD and YOLOv3 | 90.09% | - | - |
| Kutlu et al. [93] | 2020 | R-CNN | 97.52 | 88.90 | - |
| Fan et al. [94] | 2019 | ResNet50 | 98% | - | - |
| Hegde et al. [95] | 2019 | Pre-train AlexNet model | 98.9% | 98.6% | 98.7 |
| Acevedo et al. [96] | 2019 | Pre train convolutional neural network | 96.2% | - | - |
| Qin et al. [97] | 2018 | Deep Residual Learning | 76.84% | - | - |
| Tiwari et al. [98] | 2018 | Double CNN model | 97% | 83% | - |
| Hung et al. [99] | 2017 | AlexNet and Fast R CNN | 72% | - | - |

| Author | Year | Model | | | |
|---|---|---|---|---|---|
| Naz S. et al. [100] | 2017 | CNN, faster R CNN | 94.71% | 95.42 | 99.27 |
| Vogado et al. [101] | 2018 | CNN model with support vector machine | 99.20% | 99.2 | - |
| Habibzadeh et al. [102] | 2018 | ResNet and Inception | 99.46% | 99.89 | - |
| Song et al. [103] | 2014 | CNN | 94.50% | 87.26 | - |
| Fatih et al. [104] | 2019 | MRMR feature selection -ELM and CNN | 97.37% | - | - |
| Rehman et al. [105] | 2018 | Deep CNN | 97.78% | - | - |
| Bani-Hani et al. [106] | 2018 | CNN with the optimized genetic method | 91% | 91 | 97 |
| Di Ruberto et al. [107] | 2020 | Pre trained AlexNet | 97.93 | 99.6 | - |
| Loey et al. [108] | 2020 | Pre trained CNN AlexNet | 100% | 100 | 98.2 |
| Ma et al. [109] | 2020 | Generative Adversarial Network and residual neural network | 91.7% | 92% | - |
| Baydilli et al. [110] | 2020 | Capsule Networks | 96.86% | 92.5 | 98.6 |
| Tobias et al. [111] | 2020 | Faster Residual Neural Network | 83.25% | - | - |
| Kassani et al. [112] | 2019 | Hybrid DL based model | 96.17% | 95.17 | 98.58 |
| N.Baghel et al. [113] | 2022 | CNN | 98.51% | 98.4 | - |
| A. Shahzad et al. [114] | 2022 | CNN | 98.44% | 99.96 | 99.98 |
| C. Cheuque et al. [115] | 2022 | Multilevel CNN | 98.4% | 98.3 | - |
| M. Hosseini et al. [116] | 2022 | Convolutional Neural Network | 97% | 94 | 98 |
| K.Ramya et al. [117] | 2022 | CNN-PSO | 99.2% | 94.56 | 98.78 |
| A.A.Khalil et al. [118] | 2022 | CNN | 98% | - | - |
| S. Sharma et al. [119] | 2022 | DenseNet121 | 98.84% | 98.85% | 99.61 |

**Table 9:** Comparative Analysis of Top-Performing Deep Learning Models (n = 8)

| Author | Year | Dataset | No of Images | Method | Accuracy |
|---|---|---|---|---|---|
| N.Baghel et al. [113] | 2022 | Private dataset | - | CNN | 98.51% |
| C. Cheuque et al. [115] | 2022 | Private dataset | - | Multilevel CNN | 98.4% |
| K.Ramya et al. [117] | 2022 | Merged LISC and BCCD | - | CNN-PSO | 99.2% |
| S. Sharma et al. [119] | 2022 | Kaggle | 12444 | DenseNet121 | 98.84% |
| Riaz et al. [134] | 2023 | Catholic University of Korea /Public dataset | 5000 | deep convolutional generative adversarial network | 99.9% |
| Mosabbir et al. [135] | 2023 | National Institutes of Health (NIH) dataset | 27,558 | CNN | 97.92% |
| Tunseem et al. [136] | 2023 | AML Cytomorphology LMU | 18,365 | CNN | 99.7%% |
| Kakumani et al. [137] | 2023 | Kaggle | 12515 | Pre-trained inception v3 | 99.76% |

Figure 4 illustrates the distribution of publications among 27 diverse countries across the globe. Notably, the majority of research papers originate from the United States and the Netherlands, reflecting the highest number of publications. This distribution showcases the varying levels of research papers across these countries, from the highest to the lowest number of publications. Figure 5 shows the different traditional machine learning and deep learning models for white blood cell classification. Some authors used traditional machine learning models including Decision Trees (DT), K-means, Naive Bayes Classifier (NBC), Nearest Neighbor Classifier (NNC), Support Vector Machine (SVM), Artificial Neural Networks (ANN), and thresholding. While others used deep learning convolutional neural networks for WBCs. Among all CNN model is used by most of the authors.

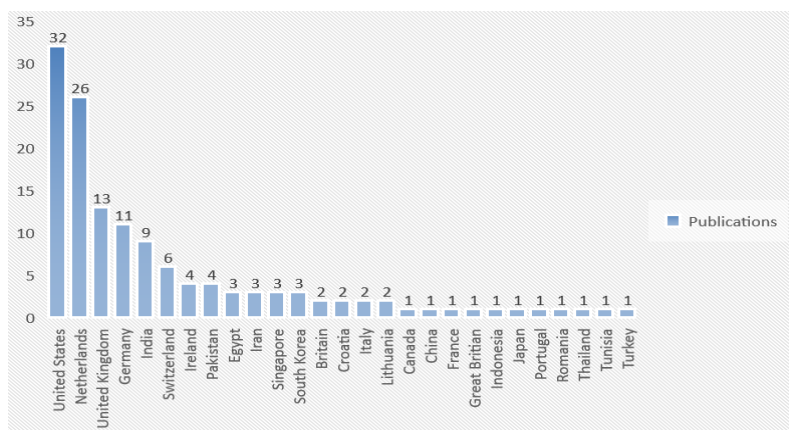

**Figure 4:** Country-wise research publications

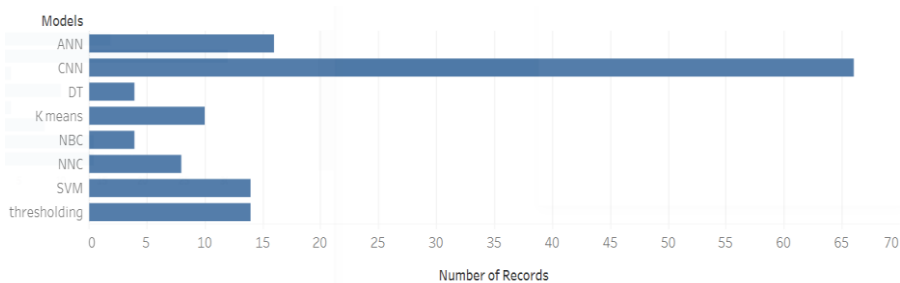

**Figure 5**. Commonly used models for White blood cells classification

## V. LIMITATIONS OF PREVIOUS STUDIES AND FUTURE CHALLENGES

ML/DL researchers have made significant advances in increasingly accurate classification of white blood cells in recent years. Among all techniques based on SVM, Sajjad et al. [68] achieved maximum accuracy, sensitivity, and specificity of 98.6%, 96.2%, and 98.5%. Using KNN, Abdeldaim et al. [75] achieved maximum accuracy of 98.6%. Similarly, using ANN, Hegde et al. [77] acquired accuracy, sensitivity, and specificity of 99%, 99.4%, and 99.18%. Using DL methods, Loey et al. achieved maximum precision and sensitivity of 100% each and specificity of 98.2%; However, while many researchers achieved close to maximum performance, several limitations and constraints have been associated with previous and current techniques. Accordingly, the research community, faces several fundamental obstacles in the field of MIA that must be accepted and resolved. These include the lack of easily accessible, large, high-quality datasets, a shortage of dedicated medical professionals, and the complexity of Transfer Learning and Deep Learning methods. Several DML strategies, mathematical and theoretical foundations are also a source of several challenges [120-121], with unsupervised or semi-supervised systems needed to address these issues [122]. Moreover, TML and DL-based MIA applications and systems still have significant work to adopt "real-time application".

### A. Lack of publicly accessible datasets

The lack of publicly accessible datasets represents the primary issue affecting medical image analysis. Scientists need to inspire health organizations to address this problem, it would be beneficial if high-quality data is available to the researchers. Initiatives promoting open data availability from various health organizations worldwide should also be encouraged. However, authorization should also be required (e.g., hospital data and conditional access to datasets). When data are readily available in large quantities, just like in other fields such as environmental science, weather forecasting, and bioinformatics, the issue becomes more relevant for research (e.g., video summarization [123], IoT [124], energy management [125], and so on). Acquiring massive, high-quality datasets with accurate labeling is crucial for MIA applications.

### B. Generalization skills for trained predictors

Another very significant challenge associated with MIA and leucocyte identification and classification is the availability of appropriately trained predictors. A perfect learning method that balances computational efficiency with generalization capacity is required to solve this issue. To build a model with impressive generalization capabilities, a learning approach that incorporates true or random labels is necessary. This approach provides efficient training algorithms and practical tools to handle available datasets using accurate or arbitrary labels. Many MIA tasks, including identifying brain tumors, lung cancer, breast cancer, and leucocytes, have shown significant empirical success. Despite the inherent challenges posed by non-convex optimization, basic techniques such as stochastic gradient descent (SGD) can efficiently discover viable solutions, effectively minimizing training errors. More interestingly, the networks created in this manner have strong generalization capabilities [126], even when there are far more parameters than training data [127]. Only reducing the training error during model training is insufficient. The choice of global minima greatly impacts the generalization behavior of the predictor. It is crucial to select the appropriate algorithm to minimize training error for better results. Different initialization, update, learning rate, and halting conditions for optimization algorithms will result in global minima with various degrees of generalizability.

### C. Reliable methods for real-world scenarios

TML and DL approaches are dependable for real-world health diagnosis systems [128], but developing accurate MIA and leucocyte classification models demands expertise and technical proficiency. In the future, researchers must investigate such accurate and reliable procedures that can be used in real-world healthcare situations without the necessity for medical specialists.

## VI. FUTURE DIRECTIONS FOR RESEARCH

The biomedical engineering and research community should dedicate substantial effort to support MIA, particularly leukocyte examination in blood images, due to the significant challenges faced by the MIA community, as detailed in section V.

### A. Data augmentation methods to complete the dataset deficit

This work addresses the issue of limited dataset availability in MIA and leucocyte classification. We present data augmentation approach and leverage transfer learning algorithms to enhance the identification of WBCs.

### B. Technical skills and medical experience required

TML and DL models have shown significant potential for computer-aided MIA-based diagnostic applications, and popular open-source frameworks like TensorFlow, Caffe, and Keras offer access to these advanced models [129]. Developing effective machine learning models for medical image analysis (MIA) requires careful consideration and expertise in the clinical and medical domains. It is essential to choose and train the suitable model to achieve accurate and reliable results in MIA applications.

### C. Resource-aware DL models for classifying leukocytes

Medical Image Analysis (MIA) with the adoption of advanced DL models like GANs, R-CNN, Fast R-CNN, and faster R-CNN, along with the integration of TML and DL methods. These models have shown superior performance in tasks like brain tumor detection, leukocyte classification, breast cancer diagnosis, and various other MIA applications. However, their biggest concerns are the significant memory needs and computational costs. Therefore, it is necessary to investigate the computationally and environmentally friendly TML and DL models for leukocyte analysis in blood images.

### D. Models for the detection and classification of leukocytes

DNNs provide a superior alternative to conventional learning techniques. The end-to-end models, especially CNNs, stems from their efficient process and the capability to classify leucocytes into five classes [130]. These models compete with complex MIA models built on DNN based on data-driven learning methodologies. WBC detection and categorization in images can also be accomplished using a variety of end-to-end designs [131-132].

### E. TML AND DL universal evaluation in MIA

The MIA research community often relies on subjective evaluation methods, which can be challenging, inefficient, and prone to errors. Therefore, comprehensive evaluation techniques that can automatically assess the effectiveness of Traditional Machine Learning (TML) and Deep Learning (DL) models for MIA from various views.

## CONCLUSION

This work provides a comprehensive review of the TML and DL techniques applied to WBCs classification. We thoroughly explored and compared various methods for WBC categorization in this context. The data for this research is compiled from 136 primary papers published between 2006 and 2023. These papers encompass TML and DL methodologies for leukocyte classification and their applications in medical diagnosis. The comprehensive analysis of these studies reveals the significant contributions of TML and DL techniques to MIA. The main objective of this work is to identify and synthesize the myriad TML and DL applications in MIA, particularly in the domain of leucocyte classification in blood smear images. This research aims to provide valuable insights into the complex characteristics of TML and DL in MIA by thoroughly analyzing existing literature. Based on literature review outcomes, Deep Learning models like CNNs for image classification and GANs for data augmentation would be used. The study's results emphasize the importance of conducting more research on using TML and DL methods effectively in MIA and classifying leucocytes in blood smear images. Besides leucocyte classification, this study explored applications for advanced DL models. Collecting all these data in this study will help the research industry by indicating where they should focus their future investigation of TML and DL models for MIA. These methods have the potential to lead to significant advancements in speech analysis, natural language processing (NLP), and medical imaging in the future. In addition to WBCs, TML and DL approaches are employed to identify and categorize various MIA domains, such as the analysis of MRI, CT, X-ray, and ultrasound images. Blood smear images are a growing field in MIA that has drawn attention from the research community over the past three decades. Additionally, we recognized the problems, instructions, and solutions for the developments of TML and DL models in MIA, notably for classifying WBCs in blood smear images. The potential of TML and DL approaches will be used to expand our research to include different MIA domains, including MRI, CT, Ultrasound, and X-ray images.